\documentclass[%
showpacs,preprintnumbers,
amsmath,amssymb,
aps,
prl,
twocolumn,
]{revtex4}

\usepackage{graphicx}
\usepackage{soul}

\newcommand{\ssm}{\scriptscriptstyle\rm}
\renewcommand{\phi}{\varphi}
\newcommand{\pdag}{\phantom{\dag}}

\def \be{\begin{equation}}
\def \ee{\end{equation}}
\def \ba{\begin{array}}
\def \ea{\end{array}}
\def \bea{\begin{eqnarray}}
\def \eea{\end{eqnarray}}

\def \W{{\Omega}}
\def \e{{\epsilon}}

\def \a{{\alpha}}

\def \b{{\beta}}
\def \g{{\gamma}}
\def \D{{\Delta}}

\def \s{{\sigma}}

\def \e{{\epsilon}}

\def \ket#1{{|#1\rangle }}

\begin{document}
\title{Universal Dephasing of Many-Body Rabi Oscillations of Atoms in One-Dimensional Traps}

\author{Sebastian~D.\ Huber}
\author{Ehud Altman}

\affiliation{Department of Condensed Matter Physics, The Weizmann
Institute of Science, Rehovot, 76100, Israel}

\date{\today}

\begin{abstract}
We study a quantum quench in a system of two coupled one-dimensional tubes
of interacting atoms. After the quench the system is out of equilibrium and
oscillates between the tubes with a frequency determined by microscopic
parameters. Despite the high energy at which the system is prepared we
find an emergent long time scale responsible for the dephasing of the
oscillations and a transition at which this time scale diverges. We show
that the universal properties of the dephasing and the transition arise
from an infrared orthogonality catastrophe. Furthermore, we show how
this universal behavior is realized in a realistic model of fermions with
attractive interactions.
\end{abstract}


\maketitle

Ultracold atomic systems provide a unique laboratory for investigating
nonequilibrium dynamics in correlated quantum matter. A common experimental
approach is to prepare the system in a simple known quantum state, which
is not an eigenstate of the system Hamiltonian, and observe the ensuing
time evolution \cite{Greiner02a,Hofferberth07,Widera08}. This is known as a
quantum quench. The near perfect isolation of the atoms from the environment
allows to concentrate on the intrinsic many-body dynamics following the
quench. The focus of this Letter is on the question whether such dynamics
can give rise to emergent structures or universal dynamical modes.

The theoretical description of quench phenomena poses a fundamental
challenge. Generically, the system is injected with a high energy
density. Consequently, the time evolution involves all energy scales. A
natural expectation in this case is that the subsequent dynamics would be
highly complex and nonuniversal: completely dominated by the microscopic
scales at short times; at longer times too complicated to capture in a
reasonable theory.

Yet there is accumulating evidence to the contrary. First, emergent
phenomena may certainly arise when the energy of the initial state is
not too large. In such cases a low energy effective theory appears to
give a correct description of the quench and bring out universal aspects
of the dynamics.  A good example is the decoherence dynamics of a pair
of one-dimensional condensates, prepared in a state with well defined
relative phase \cite{Hofferberth07,Jo07a}. The dynamics in this case can be
understood rather simply within the universal Luttinger liquid description
of the condensates \cite{Bistritzer07,Burkov07}. The same analysis holds
for the time evolution of coherence between two internal spin states in
a one dimensional condensate starting from a perfectly coherent state
\cite{Widera08}. Related theoretical questions of quench dynamics within
conformal field theories, of which the Luttinger liquid is a particular
example, are discussed in Refs. [\onlinecite{Calabrese06,Iucci09,Silva08}].

Second, and more surprisingly, there is numerical evidence for the existence
of emergent dynamical modes of very long and possibly diverging time scales,
even when the initial state in the quench is at high energy. Such behavior
was seen, for example, in the time evolution of the staggered magnetization
in a quantum quench of a spin-$1/2$ chain \cite{Barmettler09}. In a wide
regime the staggered moment displays rapid oscillations, which clearly
reflect the microscopic magnetic exchange interactions. On the other
hand, the decay of the oscillations occurs on a much longer time scale,
which even diverges on approaching the $XY$ limit of the spin chain. Such
behavior is highly suggestive of a theoretical description, which separates
out the fast dynamical modes in order to focus on the emergent long time
dynamics. We do not know, at present, of a consistent way to accomplish
such separation in the spin model of Ref. \cite{Barmettler09}.
%
\begin{figure}[b]
\includegraphics{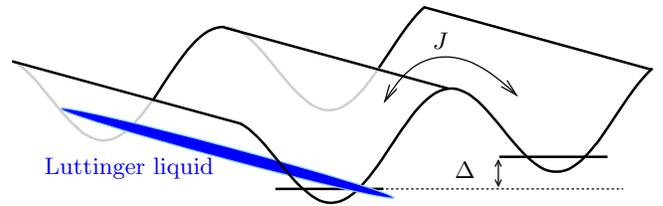}
\caption{
Two tubes detuned by $\Delta$ and coupled via a tunneling amplitude $J$. For
$|\Delta/J| \gg 1$ the ground state is given by all particles residing in
the lower tube. We investigate the behavior of the density after a sudden
ramp to $\Delta\approx0$.
}
\label{fig:setup}
\end{figure}

In this Letter we achieve separation of scales within a related model of
a quantum quench, which is relevant to experiments with one dimensional
bosons or fermions in coupled tubes, Fig.~\ref{fig:setup}. Consider a pair
of tubes, where in the initial state one of the tubes is at a much lower
potential. The ground state corresponds to the quantum liquid filling that
tube. How does this ground state evolve when the empty tube is suddenly
brought to near resonance with the filled tube?

The regime of interest to us in this work is when the tunnel coupling
between the two tubes $J$, defines the largest frequency scale in the
problem. That is, $J$ is much larger than the chemical potential $\mu$. In
this regime the problem defines a direct many-body generalization of
the usual Rabi oscillations of a single particle in a double well. The
microscopic Hamiltonian of the system is given by $H=H_0+\sum_{n=R,L}
H^{\ssm 1D}_{n}$ with
\begin{align}
\label{eqn:hamiltonian-fermion}
H_{0}&=
\sum_{\sigma}
\int \frac{dk}{2\pi}\,
\left(c_{L\sigma k}^{\dag}\,,\,c_{R\sigma k}^{\dag}\right)
\mathcal{H}_{0}
\begin{pmatrix}
c_{L\sigma k}\\
c_{R\sigma k}
\end{pmatrix}, \\
\mathcal{H}_{0} &= J\tau^x+\frac{\Delta}{2}\tau^z,\\
H^{\ssm 1D}_{n}&=
\sum_{\sigma}\int \frac{dk}{2\pi}\,
\frac{\hbar^{2}k^{2}}{2m}
c_{n\sigma k}^{\dag} c_{n\sigma k}^{\pdag}
+ g \!\int dx\, \rho_{n\uparrow}\rho_{n\downarrow}.
\end{align}
Here, $\tau^\a$ are Pauli matrices acting in the pseudo-spin ``tube-space''
and $n=L,R$ (left/right) is the tube index. $\Delta$ is the detuning between
the tubes after the quench. The Hamiltonian is written for fermions with
mass $m$ and spin $\sigma=\uparrow,\downarrow$; $\rho_{n\s}$ is the local
density operator. For bosons the Hamiltonian takes an identical form only
dropping the spin subscript. The effective contact interaction $g$ depends
on the three-dimensional scattering length $a_{s}$ and the transverse
confinement \cite{Bergeman03,Moritz05}.

In absence of interactions, under the influence of
$H_0$, all particles perform Rabi oscillations of frequency
$\W=\hbar^{-1}\sqrt{J^{2}+\Delta^{2}/4}$,  perpetually in phase with
each other. In general, when such rotating spins are coupled to a dynamic
environment, the {\em fast} time-scale $\W^{-1}$ is supplemented by {\em
slow} time-scales $T_{1(2)}$, responsible for relaxation and dephasing of
the Rabi oscillations, respectively. Here, these slow time scales emerge
due to an intrinsic mechanism. The fluctuations of the one dimensional
quantum liquid will supply the sought slow dynamics in a tractable way.

Before proceeding we outline the general strategy for solving this problem
and our main results.  In order to separate out the rapid oscillations, we
work within a reference frame rotating with the pseudo-spins (interaction
picture with respect to $H_0$). In this frame the two modes corresponding
to left and right tubes are replaced by the  filled, in-phase rotating
mode (mode 1) and the orthogonal, out-of-phase rotating mode (mode 2).
Without interactions all particles rotate perpetually in mode 1. The
interactions give rise both to an effective detuning $\D_{\ssm eff}$,
as well as particle hopping terms between the two modes.
 The particle current from the in-phase rotating mode to the out-of-phase
 mode is the dephasing rate.

For small effective detuning we can compute the dephasing rate by treating
the quantum liquid in the filled tube, within its low energy effective
theory, the Luttinger liquid.   The dephasing rate as a function
of effective detuning is  found to be a power law $T_{2}^{-1}\sim
(\D_{\ssm eff}/\mu)^\beta$,  where the exponent $\b$ can be positive or
negative and reflects the critical fluctuations of the Luttinger liquid.
A transition to positive $\beta$ corresponds to an orthogonality catastrophe
\cite{Anderson67}, whereby tunneling of particles out of the Luttinger liquid
(dephasing) is suppressed at zero detuning by the critical fluctuations. For
negative $\beta$ the tunneling ( dephasing) rate is enhanced rather than
suppressed by the fluctuations.  We show that in an experimentally relevant
system of fermions with attractive interactions, the exponent $\beta$
can be tuned from negative to positive by tuning, for example, $a_{s}$.

We now turn to derive the main results. The transformation to the rotating
frame is accomplished by
\begin{equation}
\label{eqn:rabi}
\begin{pmatrix}
|1\rangle \\
|2\rangle
\end{pmatrix}
=
e^{i\mathcal{H}_0 t/\hbar}
\begin{pmatrix}
|L\rangle\\
|R\rangle
\end{pmatrix}.
\end{equation}
In this frame $H_0$ is eliminated and the Hamiltonian is
\begin{equation}
\label{eqn:12}
{\tilde H}={\tilde H}_{1}+{\tilde H}_{2}+{\tilde H}_{12}(t).
\end{equation}
The intramode Hamiltonians ${\tilde H}_{1}$ and ${\tilde H}_{2}$ are of the
same form as $H_n^{\ssm 1D}$ in Eq.~(\ref{eqn:hamiltonian-fermion}). The
coupling between the two modes ${\tilde H}_{12}(t)$ arises because of the
interactions and contains three different contributions: a single particle
hopping $\tilde h_{\ssm 1p} \sim ga(t) \sum_{\sigma} \psi_{1\sigma}^{\dag}
\psi_{2\sigma}^{\pdag} (\rho_{1-\sigma} - \rho_{2-\sigma}) + {\rm H.c.}$;
a pair hopping term $\tilde h_{\ssm 2p} \sim gb(t) \psi_{1\uparrow}^{\dag}
\psi_{1\downarrow}^{\dag} \psi_{2\uparrow}^{\pdag} \psi_{2\downarrow}^{\pdag}
+ {\rm H.c.}$; and an inter-mode interaction $\tilde h_{\ssm int} \sim
gc(t) \sum_{\sigma} \rho_{1\sigma} \rho_{2-\sigma} + \psi_{1\sigma}^{\dag}
\psi_{1-\sigma}^{\pdag} \psi_{2-\sigma}^{\dag} \psi_{2\sigma}^{\pdag}$.
Note that if the interactions were symmetric under $\mathsf{SU}(2)$
rotations in the tube space then ${\tilde H}_{12}$ would vanish. However,
the fact that particles interact only within the same tube constitutes a
strong breaking of this $\mathsf{SU}(2)$ symmetry.

The coupling constants appearing in ${\tilde H}_{12}(t)$ are time
dependent with a periodicity corresponding to a fraction of the Rabi
period $T$. Since the Rabi frequency $\Omega$ is the highest frequency
scale in the system we can treat this time dependence within a systematic
expansion in $\mu/\Omega$ \cite{Grozdanov88, *Kapitza51}. This approach is
the quantum equivalent of the Kapitza pendulum problem \cite{Kapitza51}.
The leading order contribution is obtained by replacing $a(t)$, $b(t)$,
and $c(t)$ by their average value
\begin{equation}
\nonumber
x_{\ssm int}=
\frac{1}{T}\int_{0}^{T}\!\!dt\, x(t)=
\frac{2}{[4+\alpha^{2}]^{2}}\times
\begin{cases}
\alpha(1+\alpha^{2}) & x=a, \\
-2+\alpha^{2} & x=b, \\
-2(1+\alpha^{2}) & x=c,
\end{cases}
\label{eqn:averages}
\end{equation}
where $\alpha=\D/J$. The oscillatory time dependence of the interaction leads
to sub-leading corrections at order $\mu/\Omega$ to the calculation of $T_2$.

We focus on the effective time independent problem relevant for calculation
of $T_2$. It consists of a full tube corresponding to the particles in the
in-phase mode $\ket{1}$ and an empty tube representing the complementary
mode $\ket{2}$.  For the purpose of computing the tunneling rate, the
empty mode can be represented by
\begin{equation}
\label{eqn:effc}
\tilde{H}_{2} =
\sum_{\sigma}\int \frac{dk}{2\pi}\,
\left(\frac{\hbar^{2}k^{2}}{2m}-\mu^{\ssm eff}_{2}\right)
c_{2\sigma k}^{\dag} c_{2\sigma k}^{\pdag}.
\end{equation}
Here $\mu^{\ssm eff}_2=c_{\ssm int}g\rho_{0}/2$ is a Hartree shift due to
the effective interaction between particles in the two modes.

The full mode with density $\rho_{0}$ experiences an interaction
induced shift in its chemical potential $\mu^{\ssm eff}_1$ as well. We
discuss the exact form of this shift away from the Fermi energy
$\epsilon_{F}=\pi^{2}\hbar^{2}\rho^{2}_{0}/8m$ below. Within a microscopic
model we will also show that tuning the interactions can lead to an
effective resonance between the two modes. Namely $\D_{\ssm eff}\equiv
\mu^{\ssm eff}_2-\mu^{\ssm eff}_1=0$.

For the formulation of the long-wavelength Luttinger liquid theory near
the effective resonance we introduce the slow fermion modes $\psi_\pm$,
such that $\psi_{1\sigma}(x) \approx \psi_{+,\sigma}(x) e^{ik_{F}x} +
\psi_{-,\sigma}(x) e^{-ik_{F}x}$, and $k_{F}$ is the Fermi momentum. As
usual we express the  Hamiltonian (\ref{eqn:12}) using these fields and
neglect terms which oscillate as $e^{i n k_F x}$. In particular, for this
reason the single particle term $\propto a_{\ssm int}$ disappears from
the long-wavelength theory.

Next, we bosonize the slow fields via $\psi_{\pm,\sigma}(x) \propto
\exp\{-i [ \pm \theta_{\rho} - \phi_{\rho} + \sigma( \pm \theta_{\sigma} -
\phi_{\sigma} )] / \sqrt{2}\}$ \cite{Giamarchi04}. We shall concentrate on
the case of attractive interactions, where fermions form either (singlet)
Cooper pairs or tightly bound molecules \cite{Fuchs04}. From the point
of view of the long-wavelength theory, the fermions form a Luther-Emery
liquid with a spin gap $\D_\s$ \cite{Luther74}. The low energy degrees of
freedom therefore lie in the charge sector.
\begin{equation}
\label{eqn:effi}
\tilde{H}^\rho_{1} = \frac{\hbar v_{\rho}}{2\pi}\int dx\,
\biggl\{
K_{\rho}
[\partial_{x}\phi_{\rho}(x)]^{2}+\frac{1}{K_{\rho}}[\partial_{x}\theta_{\rho}(x)]^{2}
\biggr\}.
\end{equation}
The phonon velocity $v_{\rho}$ is related to the Fermi velocity
$v_{F}=\pi\hbar\rho_{0}/2m$ by $K_{\rho}=v_{F}/v_{\rho}$; later on we
derive an explicit expression for $K_{\rho}$ from the microscopic theory.
The bosonized form of the tunneling terms between the two modes is given by
\begin{multline}
\label{eqn:effic}
\tilde{H}_{12}=
-2g \int dx\,
\biggl(
c_{\ssm int}
\left[
\cos\bigl(\sqrt{2}\theta_{\sigma}\bigr)
e^{i\sqrt{2}\phi_{\sigma}}
\psi_{2\downarrow}^{\dag}\psi_{2\uparrow}^{\pdag}
\right]
\\
+
b_{\ssm int}
\left[
\cos\bigl(\sqrt{2}\theta_{\sigma}\bigr)
e^{-i\sqrt{2}\phi_{\rho}}
\psi_{2\downarrow}\psi_{2\uparrow}
\right]
+{\rm H.c.}
\biggr).
\end{multline}

The set of Eqs. (\ref{eqn:effc})--(\ref{eqn:effic}) allows us to
calculate the evolution of the Rabi oscillations after the quench. We
remark that due to the spin gap $\Delta_{\s}$, the first term in
(\ref{eqn:effic}) is irrelevant for $\D_{\ssm eff}<\D_\s$. In the
second term of (\ref{eqn:effic}), the spin-part of the coupling leads
to a logarithmic correction of the ``coupling constant'' $gb_{\ssm int}$
\cite{Giamarchi04}. Since we are only interested in the leading behavior
of the dephasing rate, we can focus purely on the charge sector. To second
order in the coupling $gb_{\ssm int}$, we obtain via a Fermi golden rule
calculation (FGR) the dephasing rate
\begin{multline}
\label{eqn:integral}
1/T_{2}^{F}
\sim
\int dkdqd\omega d\nu\,
A_{2}(k,\omega)
A_{2}(q,\nu)\\
\times
A_{1}(-k-q,-\omega-\nu).
\end{multline}
The spectral densities entering the FGR expression are
\begin{align}
A_{2}(k,\omega)&=
2\pi\delta(\omega-\hbar^{2}k^{2}/2m-\mu),\\
\label{eqn:a2phi}
A_{1}(k,\omega)&=
\frac{1}{\epsilon_{F}}\left(\frac{\epsilon_{F}^{2}}{\omega^{2}-\hbar^{2}v_{\rho}^{2}k^{2}}\right)^{1-\frac{1}{2K_{\rho}}}.
\end{align}
Performing the integral in (\ref{eqn:integral}) we obtain the dephasing
rate, which is one of our main results
\begin{equation}
\label{eqn:central}
1/T_{2} \sim \frac{ (b_{\ssm int}g\rho_0)^2}{\hbar\e_F}
(\Delta_{\ssm eff}/\e_F)^{-1+\frac{2}{K_{\rho}}}.
\end{equation}
%
%
\begin{figure}[t]
\includegraphics{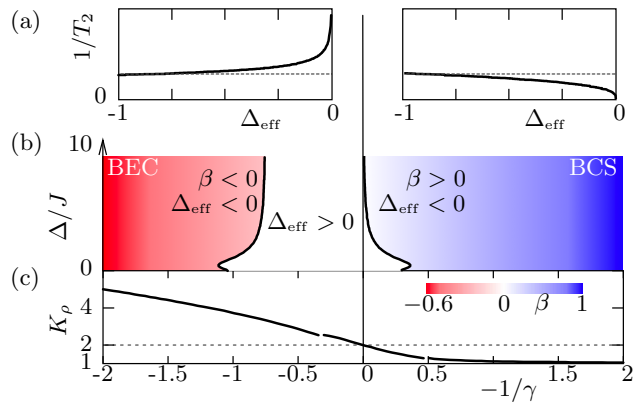}
\caption{
"Phase diagram" of dephasing rates. (a) Dephasing rate for the coherent
Rabi oscillations after a sudden quench to a fixed effective detuning
$\Delta_{\ssm eff}$ (arb. units). For $\gamma<0$ (BCS) the dephasing rate
is suppressed on the effective resonance, while for $\gamma>0$ (BEC) it
diverges. (b) Phase diagram in the plane of $1/\gamma$---$\Delta/J$. To the
left is the BEC regime, where the dephasing rate diverges, and to the right
the BCS regime where it is suppressed. The locus of effective resonance
($\D_{\ssm eff}=0$) is marked by the lines enclosing the white area. (c)
The Luttinger parameter as a function of $1/\gamma$ \cite{Fuchs04}.
}
\label{fig:pd-fermion}
\end{figure}
The most striking aspect of this result is that the point $K_{\rho}=2$
represents a transition in the nonequilibrium dynamics. While for
$K_{\rho}>2$, the Rabi-oscillations are over-damped, for $K_{\rho}<2$ the
dephasing rate is suppressed as a power-law as $\D_{\ssm eff}\to 0$. That
is, the time scale $T_2$ diverges [see Fig.~\ref{fig:pd-fermion}(a)]. The
universal behavior of $T_2$ is generated by an infra-red singularity,
or orthogonality catastrophe, due to coupling with critical charge
fluctuations in the Luttinger liquid.  The effect is analogous to blocking
of tunneling through a point contact between one dimensional electron
liquids \cite{Kane92b}. Scaling analysis, similar to Ref. \cite{Kane92b},
shows that coupling between the two rotating modes is irrelevant at all
orders. Therefore absence of dephasing in the limit $\D_{\ssm eff}\to 0$
is valid beyond the FGR analysis.

All this is strictly true only within our static (time average)
approximation. The neglected rapid oscillation of the coupling in
the rotating frame will inevitably lead to some dephasing, as will
inhomogeneities and finite temperature. In the opposite regime of $K>2$,
the divergence of the dephasing rate signals failure of FGR as $\D_{\ssm
eff}\to 0$ and it will be cut off in reality. The practical implications
however remain unchanged: overdamped oscillations for $K>2$ versus strongly
suppressed dephasing for $K<2$ as $\D_{\ssm eff}\to 0$.

It is interesting to contrast the dephasing of oscillations in the
interacting quantum liquid with the usual problem of decoupled two-level
systems connected to a reservoir. In both cases ``$T_2$'' processes
correspond to flipping a two-level system from in-phase to out-of-phase
precession. Contrary to the usual case, here the in- and out-of-phase modes
are not strictly degenerate. Transferring a particle into the out of phase
mode can involve a change in the interaction or axial kinetic energy. The
excess energy is absorbed by excitations of the quantum liquid.

In the following we show how a low energy theory with $K_{\rho} \in
[1,\infty]$, $\Delta_{s}>0$, and $\Delta_{\ssm eff} \rightarrow 0 $
is realized for fermions confined to quasi-one-dimension. We follow
the analysis of Fuchs {\em et al.} \cite{Fuchs04}, who discussed an
exactly solvable model relevant to the BCS to BEC cross-over of quasi
one-dimensional fermions. The model is parametrized by a dimensionless
interaction constant $\gamma=g/(\hbar^{2}\rho_{0})$, where $g$ is the
coupling constant of the Hamiltonian (\ref{eqn:hamiltonian-fermion}). For
$1/\gamma \rightarrow-\infty$ the system is in the weakly interacting BCS
regime with a spin gap $\Delta_{\s}\propto \exp[-\pi^{2}/2|\gamma|]$. For
$1/\gamma>0$ the Hamiltonian (\ref{eqn:hamiltonian-fermion}) does not
support a bound state. However, this is an artifact of the strictly single
channel description. In reality the bound state, and the spin gap go up
to the transverse trap frequency as $1/\g\to 0^+$  violating the single
channel assumption, this is the so called confinement induced resonance
(CIR) \cite{Bergeman03,Moritz05}. Fuchs {\em et al.} therefore considered
the model (\ref{eqn:hamiltonian-fermion}) supplemented with a bound state
at $\epsilon_{b}=-\infty$ (or $\Delta_{\s}=\infty$) for $\gamma>0$. We can
extract the relevant parameters for the low energy theory from their exact
solution. In the   crossover regime $1/\gamma \approx 0$, the Luttinger
parameter $K_{\rho}^{-1} \approx (1-1/\gamma)/2$ and the chemical
potential $\mu^{\ssm eff}_{1} \approx \epsilon_{F}(1/4-1/3\gamma)$. It
is interesting to note that the transition between over-damped Rabi
oscillations to suppressed dephasing for $K_\rho<2$ occurs exactly at the
CIR point $1/\g=0$, see Fig.~\ref{fig:pd-fermion}(c).

To find the location of the effective resonance $\D_{\ssm eff}=0$, we
compute the interaction-induced detuning \cite{Fuchs04}
\begin{equation}
\Delta_{\ssm eff}
=\mu^{\ssm eff}_{1}-\mu^{\ssm eff}_{2}
\approx
\epsilon_{F}
\left[
\frac{1}{4}+\frac{1}{3|\gamma|}-|\gamma c_{\ssm int}| \frac{4}{\pi^{2}}
\right].
\end{equation}
We see that $\Delta_{\ssm eff} = 0$ requires $1/\gamma \lesssim 1$, which
is in the strongly interacting, or cross-over regime. The need for strong
interactions is not surprising because the interaction effects have to
overcome the Fermi energy of the full mode. In Fig.~\ref{fig:pd-fermion}(b)
we show the locus of the effective resonance on the two sides of
the ``transition'' in the plane of bare detuning versus interaction
parameter. The very different trends on the two sides of the BCS to BEC
crossover  can be observed even if one does not tune exactly to $\D_{\ssm
eff}=0$. The white region in Fig.~\ref{fig:pd-fermion}(b) corresponds
to $\D_{\ssm eff}>0$ where within our FGR calculation no particles are
transferred out of the full mode.

We now turn to briefly discuss a similar setup of spinless bosons with
repulsive interactions. As mentioned, this situation is described
by Eq.~(\ref{eqn:hamiltonian-fermion}) dropping the spin index. The
transformation to a rotating frame remains unchanged. Because of the absence
of a Fermi energy the effective detuning is different, and in the Hartree
approximation given by $\Delta{\ssm eff}=\rho_{0} g [(1-|c_{\ssm int}|) -
2|c_{\ssm int}|]$.  The low energy theory of the interacting bosons is also
a Luttinger liquid. However, in contrast to the fermion case, the single
particle tunneling does not vanish on going to the long wavelength limit. In
fact it is the most relevant term leading to $1/T_{2} \sim \Delta_{\ssm
eff}^{-1+\frac{1}{2K}}$. Here, $K$ is the Luttinger parameter controlling
the decay of correlations in the filled mode. An orthogonality catastrophe,
similar to the Fermions, occurs for $K<1/2$, which requires bosons with
longer range (e.g. dipolar) interactions.  Alternatively in the presence of
a commensurate lattice potential the transition to a gapped Mott insulator
would be accompanied by a transition to suppressed dephasing.

In conclusion, ultracold atoms oscillating between a pair of one
dimensional traps following a quantum quench constitute a natural many-body
generalization of Rabi oscillations. We described universal behavior in
the dephasing rate seen as power-law dependence on an effective detuning
parameter. A transition from enhanced to suppressed dephasing at zero
detuning is found below a critical value of the Luttinger parameter,
which controls the correlation decay exponent in the quantum liquid.
The transition is generated by an orthogonality catastrophe in the
coupling to critical one dimensional fluctuations. We showed that such
a non-equilibrium transition can be observed in a system of attractive
fermions in the BCS to BEC cross-over.

We acknowledge stimulating discussions with I. Bloch, S. Trotzky, and
E. Demler. This work was supported by DIP, ISF, and the US-Israel BSF.

\end{document}